\documentstyle[aps]{revtex}
\draft
\begin{document}

\twocolumn

\title{The bending of light and the cosmological constant}
\author{Kayll Lake}
\address{ Department of Physics, Queen's University, Kingston, Ontario, Canada, K7L 3N6}

\date{\today}
\maketitle
\begin{abstract}
The bending of light in Kottler space (the Schwarzschild vacuum
with cosmological constant) is examined. Unlike the advance of the
perihelion, the cosmological constant produces no change in the
bending of light. In this note we examine the conditions under
which this statement holds.
\end{abstract}

\vskip2pc

\section{Introduction}
The deflection of light is one of the ``classical" tests of
general relativity \cite{will}. With the continuing interest in
the cosmological constant \cite{cos} one would assume that the
effect of a cosmological constant on the deflection of light is
well known. Certainly the effect of $\Lambda$ on the advance of
the perihelion is well known \cite{rindler}. This appears not to
be the case and is the subject of this note.

Gravitational lensing is now a common tool in astrophysics. There
are two approaches to gravitational lensing: the thin lens
approximation and the exact approach. These differ fundamentally
in that in the thin lens approach there is a background / lens
split whereas in the exact approach one uses the full null
geodesic equations in an exact solution of Einstein's equations.
In a cosmological context, even for strong-lensing calculations,
some variant of the thin lens approximation is adequate and used
exclusively. For the purposes of the present calculation, however,
only the exact approach is appropriate. Recent studies of lensing
in the Schwarzschild field are available. Virbhadra and Ellis
\cite{virb} have examined a strong-field version of the thin lens
approximation and Newman  and coworkers \cite{newman} have
compared the exact approach with various approximations.

 The present note involves exact lensing in Kottler
\cite{kottler} space, the Schwarzschild vacuum with a cosmological
constant ($\Lambda$). In terms of familiar curvature coordinates
$(\textsf{r},\theta,\phi,t)$ the line element is given by
\begin{equation}
\label{vacuum}
ds^2 = \frac{d\textsf{r}^2}{f(\textsf{r})} + \textsf{r}^2(d\theta^2 + \sin^2 \theta  d\phi^2)-f(\textsf{r})dt^2, \label{line}
\end{equation}
where
\begin{equation}
\label{f}
f(\textsf{r}) = 1-\frac{2 m}{\textsf{r}}-\frac{\Lambda \textsf{r}^2}{3}. \label{factor}
\end{equation}
The associated generalization of the Birkhoff theorem is well
known \cite{bonnor}. It is interesting to note that the $\Lambda$
generalization of the black hole uniqueness theorems is not known
\cite{israel}. Geodesically complete forms of the metric
(\ref{vacuum}) along with Penrose - Carter diagrams are now well
know \cite{lake2}.

\section{Null geodesics}

The coordinates $(\textsf{r},\theta,\phi,t)$ are adapted to two
Killing vectors and so geodesics of the metric (\ref{vacuum}) have
two constants of motion. The orbits are stably planar and we
choose the plane to be $\theta = \pi/2$. The momentum conjugate to
$\phi$ is the orbital angular momentum  \textit{l},
$\textsf{r}^2\dot{\phi} = \textit{l}$, and the momentum conjugate
to $t$ is the energy $\gamma$, $f(\textsf{r})\dot{t} = \gamma$
where $^. = \frac{d}{d\lambda}$ and $\lambda$ is an affine
parameter. Since we are interested here only in null geodesics, it
is convenient to reparametrize them with $\overline{\lambda}
\equiv l \lambda$ so that
\begin{equation}
\textsf{r}^2\dot{\phi} = 1, \label{lrev}
\end{equation}
and
\begin{equation}
f(\textsf{r})\dot{t} = \frac{1}{b}, \label{energyrev}
\end{equation}
where $b \equiv \frac{l}{\gamma}$. If $\Lambda = 0$ then $b$ is the impact parameter. From (\ref{line}),
(\ref{lrev}) and (\ref{energyrev}) it follows that
\begin{equation}
(\frac{d\textsf{r}}{d\overline{\lambda}})^2 =
\frac{1}{b^2}-\frac{f(\textsf{r})}{\textsf{r}^2} \label{potential}
\end{equation}
 where $\textsf{r}^2_{\Sigma} \equiv b^2 f(\textsf{r}_{\Sigma})$ defines the turning points in
$\textsf{r}$ \cite{turning}. If (\ref{vacuum}) is considered
derived from a source then $r_{\Sigma}$ is the boundary of that
source (see below). From (\ref{lrev}) and (\ref{potential}) we
have
\begin{equation}
(\frac{d u}{d \phi})^2 =(\frac{m}{b})^2+\frac{\Lambda
m^2}{3}-u^2+2u^3 \label{ulambda}
\end{equation}
where $u \equiv \frac{m}{\textsf{r}}$. Differentiation of
(\ref{ulambda}) with respect to $\phi$ of course eliminates
$\Lambda$, but this does not prove that $\Lambda$ has no effect
since (\ref{ulambda}) is the first integral of the motion
\cite{islam}. In the usual way \cite{weinberg} (\ref{ulambda})
yields, in first order, the deflection angle
\begin{equation}
\delta=4\sqrt{(\frac{m}{b})^2+\frac{\Lambda m^2}{3}}.
\label{angle}
\end{equation}
Consider the cases $\Lambda = 0$ and $\Lambda \neq 0$ separately
and assume that $r_{\Sigma}$ and $m$ are fixed, assumptions we
examine below. With these conditions it is necessary to
distinguish two values of $b$ and $\delta$, say $(b,\delta)$ for
$\Lambda=0$ and $(\overline{b},\overline{\delta})$ for
$\Lambda\neq0$. Note that from (\ref{angle}), since $\delta$ is
measurable, if $m$ is known, then $b$ is measurable, but
$\overline{b}$ is not since we do not know $\Lambda$ a priori.
From the definitions of $r_{\Sigma}$, $b$ and $\overline{b}$ we
have
\begin{equation}
\Lambda = 3 \frac{b^2-\overline{b}^2}{b^2 \overline{b}^2},
\label{lambda}
\end{equation}
so that from (\ref{angle})
\begin{equation}
\delta = \overline{\delta}. \label{lambdaeq}
\end{equation}
The fact that (\ref{lambdaeq}) is an \textit{exact} relationship
follows by rewriting (\ref{ulambda}) in the form
\begin{equation}
(\frac{d u}{d \phi})^2 =u^2_{\Sigma}-2u^3_{\Sigma}-u^2+2u^3
\label{u3}
\end{equation}
where $u_{\Sigma} \equiv \frac{m}{\textsf{r}_{\Sigma}}$. Since
$u_{\Sigma}$ is fixed by our assumptions on $r_{\Sigma}$ and $m$
it follows that (\ref{lambdaeq}) is exact. Although (\ref{u3}) can
be solved exactly (the solution is recorded in the Appendix) what
is of interest here is the effect of $\Lambda$. In the case of a
black hole, without considerations beyond null geodesics,
$u_{\Sigma}$ has to be considered merely as a parameter $<
\frac{1}{3}$. However, if one inquires into the source of
(\ref{line}) a rather more detailed analysis is required as then
both $r_{\Sigma}$ and $m$ enter as derived quantities.

\section{Source}
To see how both $r_{\Sigma}$ and $m$ enter as derived quantities,
consider the source of the external field (\ref{line}) to be a
non-singular static perfect fluid \cite{fluid}. The line element
in conventional form is (e.g., \cite{MTW})
\begin{equation}
ds^2=\frac{dr^2}{1-\frac{2M(r)}{r}}+r^2(d\theta^2+sin(\theta)^2d\phi^2)-e^{2\Phi(r)}dt^2 \label{standard}
\end{equation}
with the coordinates comoving in the sense that the fluid
streamlines are given by $u^{a}=e^{- \Phi(r)}\delta^{a}_{t}$. Note
that we have written (\ref{standard}) without $\Lambda$ motivated
by the fact that $M(r)$, on a purely geometrical basis, represents
the gravitational energy at a primitive level prior to the
introduction of Einstein's equations \cite{hayward}. This choice
is, however, merely notation.
\subsection{Generalized T-OV equation}
In terms of the perfect fluid decomposition
($T^{a}_{b}=(\rho(r)+p(r))u^{a}u_{b}+p(r)\delta^{a}_{b} -
\Lambda\delta^{a}_{b}/(8 \pi) $), solving for $\Phi'(r)$ from the
$r$-component of the conservation equations and Einstein's
equations  ($\nabla_{a}T^{a}_{r}=0$ and $G^{r}_{r}-8\pi
p(r)+\Lambda=0$) we obtain the generalized Tolman \cite{tolman1}
-Oppenheimer-Volkoff \cite{ov} (T-OV) equation
\begin{equation}
\Phi'(r)=\frac{-p'(r)}{\rho(r)+p(r)}=\frac{M(r)+4\pi p(r)r^3 - \Lambda r^3/2}{r(r-2M(r))}, \label{tov}
\end{equation}
where, from the $t$ component of the Einstein equations ($G^{t}_{t}=-8\pi \rho(r)-\Lambda$),
\begin{equation}
4\pi \rho(r)+\Lambda/2=\frac{M'(r)}{r^2}. \label{rho}
\end{equation}
Despite that fact that the T-OV equation has been known for over
sixty years, only recently \cite{br} has its mathematical
structure been fully appreciated, even for $\Lambda=0$.

\subsection{Junction Conditions}

The junction of a static perfect fluid onto vacuum in spherical
symmetry by way of the Darmois - Israel conditions is well
understood \cite{MTW}. Here we follow Musgrave and Lake \cite{ML}.
To summarize, the continuity of the first fundamental form
(intrinsic metric) associated with the boundary ($\Sigma$) ensures
that the continuity of $\theta$ and $\phi$ in metrics (\ref{line})
and (\ref{standard}) is allowed and that the history of the
boundary is given by
\begin{equation}
r_{\Sigma} = \textsf{r}_{\Sigma}. \label{history}
\end{equation}
In terms of intrinsic coordinates $(\tau,\theta,\phi)$, the
continuity of the extrinsic curvature component $K_{\tau \tau}$
along with the T-OV equation (\ref{tov}) gives
\begin{equation}
p(r_{\Sigma})=0. \label{psurf}
\end{equation}
Equation (\ref{psurf}) defines $r_{\Sigma}$ and clearly $\Lambda$
plays no role in that definition. The continuity of the extrinsic
curvature components $K_{\theta \theta}$ and $K_{\phi \phi}$ give
\begin{equation}
\frac{M(r_{\Sigma})}{r_{\Sigma}} = \frac{m}{\textsf{r}_{\Sigma}}+\Lambda \textsf{r}^2_{\Sigma}/6. \label{msurf}
\end{equation}
It follows from equations (\ref{rho}) and (\ref{msurf}) that
\begin{equation}
m= \int_{0}^{r_{\Sigma}} 4 \pi r^2 \rho(r) dr \label{m}
\end{equation}
independent of the value of $\Lambda$.
\section{Discussion}
Equations (\ref{history}) and (\ref{psurf}) show that the turning
point (at the boundary of the configuration)
$\textsf{r}_{\Sigma}=r_{\Sigma}$ is unaffected by $\Lambda$. With
equation (\ref{m}) then we conclude that if the
Kottler-Schwarzschild field is considered generated by a source of
specified energy density and isotropic pressure $(\rho(r),p(r))$
then $u_{\Sigma}$ is fixed and by virtue of (\ref{u3}) $\Lambda$
has no effect on the bending of light. Clearly, it is the use of
$m$ as opposed to $M(r_{\Sigma})$ that is crucial here. When we
say ``the mass" it is $m$, and not $M(r_{\Sigma})$, that we refer
to. For example, for timelike orbits in the weak field
approximation, $m$ is ``the mass", not $M(r_{\Sigma})$. Indeed,
$M(r_{\Sigma})$, though invariantly defined, cannot be given
without a priori knowledge of $\Lambda$. With the choice that
$M(r_{\Sigma})$ alone specifies the configuration, it follows from
equation (\ref{msurf}) that $\Lambda$ would alter the bending of
light, but only by way of one's choice as to how the
Kottler-Schwarzschild field was specified. The results of this
paper can be understood in a more general context. For an
irrotational null geodesic congruence there are two types of
``focusing": Ricci focusing and Weyl focusing  \cite{Sachs}. It is
well know, and easy to show, that $\Lambda$ has no effect on Ricci
focusing. For the case studied here it is easy to see that
$\Lambda$ has no effect on the Weyl focusing either since for the
metric (\ref{vacuum}) with (\ref{f}) $\Lambda$ does not enter the
$C_{\alpha \beta}^{\:\:\:\: \gamma \delta}$ components of the Weyl
tensor.

\subsection*{Acknowledgments}
The comments of a referee helped to improve the presentation. This
work was supported by a grant from the Natural Sciences and
Engineering research Council of Canada.

\begin{center} \textbf{Appendix} \end{center}

The exact solution to (\ref{u3}) is given, up to sign, by
\begin{equation}
\phi(u) = \sqrt{\frac{k^2(u)-\Theta}{l(u)B}}F(2\sqrt{\frac{u_{\Sigma}-u}{A}},\sqrt{\frac{A}{B}}),
\end{equation}
where,
\begin{equation}
k(u) = 4u-1+2u_{\Sigma},
\end{equation}
\begin{equation}
l(u) = -2u_{\Sigma}^2+u_{\Sigma}-2u_{\Sigma}u+u-2u^2,
\end{equation}
\begin{equation}
A=6u_{\Sigma}-1+\sqrt{\Theta},
\end{equation}
\begin{equation}
B=6u_{\Sigma}-1-\sqrt{\Theta},
\end{equation}
\begin{equation}
\Theta=(1-2u_{\Sigma})(1+6u_{\Sigma}),
\end{equation}
and $F$ is the incomplete elliptic integral of the first kind.


\begin{references}
\bibitem{turning} We assume that turning points exist. This
amounts to the assumption that $3 m < \frac{1}{\sqrt{\Lambda}}$
for $\Lambda >0$ and limits the region of consideration to
$f(\textsf{r})>0$ in general.
\end{references}
\end{document}